# Undoped Strained Ge Quantum Well with Ultrahigh Mobility Grown by Reduce Pressure Chemical Vapor Deposition

Zhenzhen Kong[1,2#], Zonghu Li[3#], Gang Cao [3,4], Jiale Su [1], Yiwen Zhang[1,2], Jinbiao Liu[1,2], Jingxiong Liu[1,2], Yuhui Ren[1,2], Laiming Wei[5], Guoping Guo[3,4,6], Yuanyuan Wu[7], Henry H. Radamson[1,7], Junfeng Li[1], Zhenhua Wu[1,2], Haiou Li[3,4], Jiecheng Yang[3], Chao Zhao[8], Guilei Wang[1,4,8] ✉

[1] Integrated Circuit Advanced Process R&D Center, Institute of Microelectronics, Chinese Academy of Sciences, Beijing 100029, P. R. China;

[2] School of Integrated Circuits, University of Chinese Academy of Sciences, Beijing 100049, P. R. China;

[3] CAS Key Laboratory of Quantum Information, University of Science and Technology of China, Hefei 230026, P. R. China

[4] Hefei National Laboratory, Hefei 230088, P. R. China

[5] School of Advanced Manufacturing Engineering, Hefei University, Hefei 230601, P. R. China

[6] Origin Quantum Computing Company Limited, Hefei 230026, P. R. China

[7] Research and Development Center of Optoelectronic Hybrid IC, Guangdong Greater Bay Area Institute of Integrated Circuit and System, Guangzhou 510535, P. R. China;

[8] Beijing Superstring Academy of Memory Technology, 100176, P. R. China

*Correspondence: wangguilei@ime.ac.cn (G.W.); Tel.: +86-010-8299-5793 (G.W.);

# Equal contribution to the first author

**Abstract**

We fabricate an undoped Ge quantum well under a 30 nm relaxed $Ge_{0.8}Si_{0.2}$ shallow barrier. The bottom barrier contains $Ge_{0.8}Si_{0.2}$ (650 ℃) followed by $Ge_{0.9}Si_{0.1}$ (800 ℃) such that variation of Ge content forms a sharp interface that can suppress the threading dislocation density (TDD) penetrating into the undoped Ge quantum well. The $Ge_{0.8}Si_{0.2}$ barrier introduces enough in-plane parallel strain ($\varepsilon_{\parallel}$ strain -0.41%) in the Ge quantum well. The heterostructure field-effect transistors with a shallow buried channel obtain an ultrahigh two-dimensional hole gas (2DHG) mobility over $2\times10^6$ cm²/Vs and a very low percolation of $5.625 \times 10^{10} cm^{-2}$. A tunable fractional quantum Hall effect at high densities and high magnetic fields has also been observed. This approach defines strained germanium as providing the material basis for tuning the spin-orbit coupling strength for fast and coherent quantum computation.

## 1. Introduction

In the past few decades, due to the strong spin-orbit coupling interaction (SOI)[1, 2, 3, 4, 5, 6, 7] and weak hyperfine interaction (HFI)[8, 9, 10], holes in Ge are considered good candidates for semiconductor quantum computing[11,12]. Recently, significant progress has been achieved, such as high-fidelity single-qubit manipulation and readout, ultrafast two-qubit logic[13], and four-qubit quantum processors and their algorithms[14, 15, 16]. At the same time, the coupling between the spin qubit and resonator has also been preliminarily explored[12, 17, 18, 19]. Thus, holes in Ge provide a good platform for the large-scale expansion of semiconductor quantum computing. Meanwhile, the improvement of high-quality Ge hole materials, which will affect the performance of spin qubits, is accompanied by the development of hole spin qubits[20, 21, 22].

Strained Ge quantum well heterojunctions with high mobility are an effective material platform for realizing hole spin qubits. In early studies, modulated doping techniques were used to introduce more free activity carriers to the active layer by doping in the nonadjacent layer. A carrier mobility of 1.5 million cm$^2$/Vs can be obtained by the modulation doping method[23]. However, the strong charge noise in modulated doping has led researchers to shift to undoped GeSi/Ge heterojunctions [24]. In contrast, intrinsic material platforms can reduce leakage, the parasitic channel effect and charge noise, which are more suitable for the manipulation of quantum dots and spin qubits[18-25]. On the one hand, to obtain a high mobility in the platform, introducing biaxial compressional strain of germanium is essential. We need to prepare relaxed GeSi as a barrier layer to enhance the strain on the Ge QW. To achieve this goal, researchers usually grow Ge virtual substrate on Si substrate and then obtain relaxed GeSi barrier layer by high-temperature inverse gradient method[25]. On the other hand, there is a 4.2% lattice mismatch between silicon and germanium, which produces a large number of penetration dislocations. It is imperative to reduce the scattering introduced by dislocations and defects during growth. In previous studies, various methods were used to reduce penetration dislocations and their effects, such as a two-step deposition process, cyclic thermal annealing technology, and silicon and germanium alloy reverse gradient layer technology. In this research, we demonstrate a high-quality undoped 2D Ge/GeSi heterostructure by introducing the innovative abrupt junction $Ge_{0.9}Si_{0.1}/Ge_{0.8}Si_{0.2}$ interface at the end of the reverse grading layer, which cuts off the penetration dislocation just below that interface. In addition to inhibiting the effect of penetration defects, the steepness of GeSi barrier composition and thickness changes is ensured. The Ge quantum well was subjected to -0.41% in-plane parallel strain compressive strain by the $Ge_{0.8}Si_{0.2}$ barrier. Then, we fabricate Hall-bar shaped heterostructure field effect transistors (H-FETs) with a two-dimensional hole gas (2DHG) heterojunction under only a 32 nm $Ge_{0.8}Si_{0.2}$ barrier of the surface and use standard four-probe low-frequency lock in techniques for mobility-density and magnetotransport characterization at T = 16 mK to 766 mK. We obtained a high hole mobility over two million cm$^2$/Vs and a low percolation density of $5.625 \times 10^{10} cm^{-2}$. At the same time, a very obvious fractional quantum Hall effect was observed, which fully proves the low disorder and high quality of our material platform.

## 2. Results and discussion

### 2.1 Heterostructure Growth and Characterization

The $Ge_{0.8}Si_{0.2}$/Ge heterostructure is grown on a 200 mm Si (001) wafer by reduced pressure chemical vapor deposition (RPCVD) of ASM Epsilon 2000. The precursor used was germane (GeH$_4$) and dichlorosilane (SiH$_2$Cl$_2$). It contains 1.7 μm Ge as a virtual substrate (Ge VS), which uses a two-step method. It includes a low temperature of 450℃ and a cap of layers at 650℃. The front layer is two-dimensional flat growing and highly dislocated. Meanwhile, the second one has remarkably higher epitaxial quality. An annealing at 820°C for 20 min is followed to ensure that the defect density is minimized to a level of $10^7$ cm$^{-2}$. A 750 nm reverse grading layer with Ge contents changed from a pure Ge layer to a 90% GeSi layer, and a 200 nm Ge$_{0.9}$Si$_{0.1}$ step layer was grown at 800°C with a thinner reverse gradient buffer to obtain a better material quality. A 360 nm Ge$_{0.8}$Si$_{0.2}$ layer under barrier growth at 650°C is followed by a Ge$_{0.9}$Si$_{0.1}$ step layer, in which the sharp interface introduced by compositional mutation cuts off the TDDs that pass through the Ge quantum well. A 16 nm Ge quantum well under a 32 nm thick top Ge$_{0.8}$Si$_{0.2}$ barrier obtains enough compressive strain with the Ge$_{0.8}$Si$_{0.2}$ barrier. A thin sacrificial Si cap layer of 1.82 nm is on the surface to obtain a better interface state of the gate dielectric. The top three layers are grown at 500°C.

The schematics are shown in Figure 1a. Figure 1b shows the HRTEM image of the Ge/GeSi heterostructure. We can observe that most of the dislocations are restricted under the low-temperature Ge$_{0.8}$Si$_{0.2}$ (650°C) layer and the initial position of the Ge virtual substrate. Figure 1c is the STEM image of the Ge quantum well, which shows an absence of dislocation in the Ge quantum well. Two EDS lines of the Ge/Si components on the STEM image. The yellow line represents the concentration distribution of Si at the heterogeneous junction position, and the blue line represents Ge. We can observe that the interface between GeSi and Ge QW is clearer; however, the interface between Ge and GeSi regions is a slightly slower gradient compared to the lower interface from STEM and EDX lines due to the segregation of Ge atoms under thermodynamic driving force during initial growth[26]. The interface between Ge/GeSi can be further optimized to better limit the carrier in the quantum well. A 1.85 nm Si cap layer is grown to possibly achieve a dielectric interface superior to what GeSi could offer (Fig. 1d). As can be determined from the EDS mapping in Figure 1e, no diffusion of Si exists in the Ge quantum well, and the O element does not diffuse into the GeSi/Ge layers. The entire quantum well and the upper and lower barrier layers maintain a pure material distribution.

The strain in the quantum well position is characterized using NBD. The interplanar distances $d_1$ (200), $d_2$ (111), and $d_3$ (220) are extracted to be 2.82 Å, 3.25 Å, and 1.99 Å, respectively, corresponding to the parameters of the standard Ge crystal to obtain the three-phase compressive strains of 0.28%, 0.32% and 0.51%. At the same time, we also tested the NBD strain on the Ge buffer layer, which received 0.97% tensile strain. It is preliminarily estimated that the induced source of this tensile strain is the rapid contraction of the Ge lattice due to rapid cooling during growth because the thermal expansion coefficient of the Ge material is much greater than that of Si[25].

The roughness of the surface tested by atomic force microscopy (AFM) is 1.53 nm, and the RMS diagram shows a very inerratic surface morphology. The cross-hatch is distributed in the <110> crystal direction, and no threading dislocation density is found on the surface. This is the same as the cross-hatch morphology measured after deposition of the Ge VS and RG GeSi buffer layers. This is due to the Ge material acting as a face-centered cubic structure, arranging the atoms most closely in the <110> crystal direction. In most cases, the direction of the dislocation slip of the paradigm deformation is the closest direction along the atomic arrangement, such as the <110> crystal direction of the face core cube structure and the <111> crystal direction of the body core cube. The strain is released gradually and regularly through the dislocation slip during our growth process, and no penetrating threading dislocation that leads to dislocation scattering in Ge QW is generated.

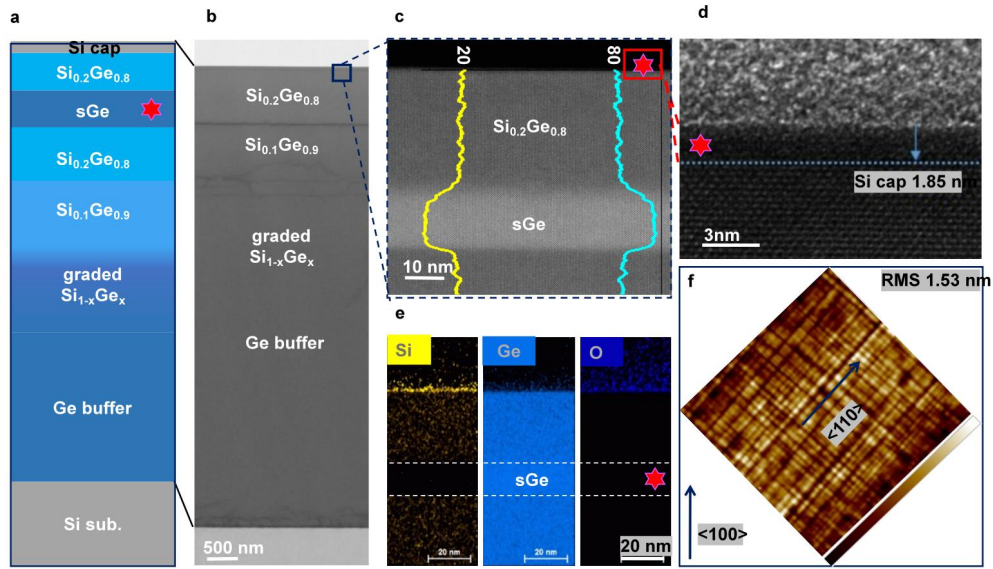

**Fig. 1 Structural characterization of a Ge/SiGe heterostructure. a** Ge/SiGe layer schematics. **b** STEM of Ge/SiGe heterostructure. **c** Ge quantum well with Ge (blue) and Si (yellow) concentration profiles by STEM/EDX. **d** HRTEM of Ge quantum well. **e** EDX mapping of Ge (blue), Si (yellow) and O (dark blue) distribution. **f** AFM of Ge/SiGe heterostructure.

In order to further determine the quality of Ge/GeSi heterostructure, high-resolution X-ray diffraction (HRXRD) and high-resolution reciprocal lattice maps (HRRLMs) were performed. As shown in Figure 2a, the Ge VS and Si substrate are fully relaxed and $Ge_{0.9}Si_{0.1}$ and $Ge_{0.8}Si_{0.2}$ barriers peak is obviously found in （004）rocking curves. However, it's hard to define sGe QW signal by （004）omega-2theta scanning. While in HRRLMs (113), the position of sGe is almost aligned to GeSi barriers peak along vertical axis in the reciprocal space showing minor strain relaxation in epi-layers (Fig. 2b).

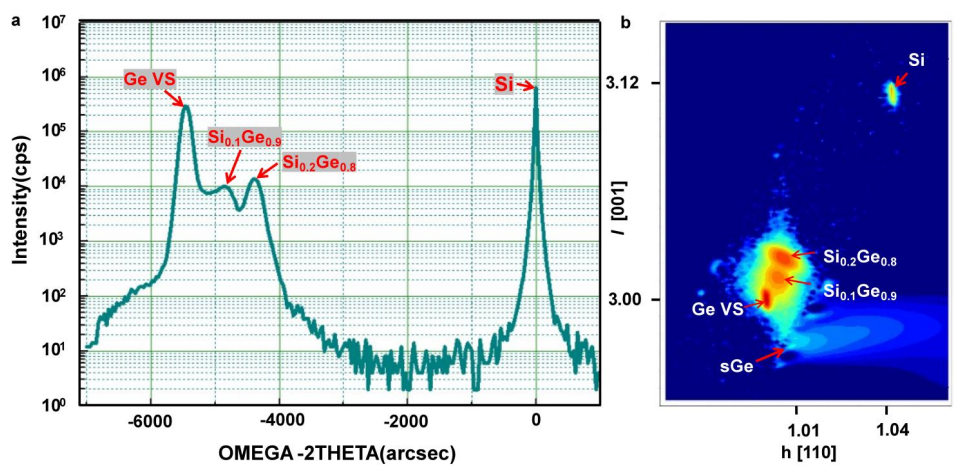

**Fig. 2 HRXRD analysis of of Ge/GeSi heterostructure. a** (004) rocking curves and **b** HRRLMs around (113) reflections.

## 2.2 Hall device preparation and characterization

To further demonstrate the growth quality of wafers, we fabricated Hall-bar shaped heterostructure field effect transistors oriented along the <110> crystallographic directions and characterized magneto transport properties at low temperature T=16 mK[3]. The device preparation process of the Hall bar is shown in Figure 3a. The device structure is shown in Fig. 3b. Figure 3c-e shows the structure of the Hall bar devices and the material distribution. Al is deposited as the Ohmic contact material. After forming the ohmic contact (300°C for 1-2 h), the quantum well does not deform or bend due to alloy annealing, and the interface is always clear and neat. This is also one of the reasons why we can obtain high mobility at low temperatures.

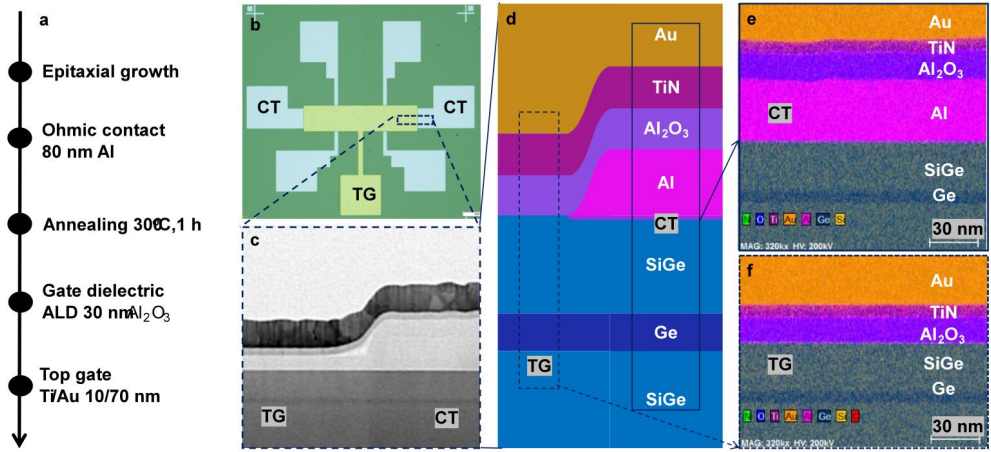

**Fig. 3 Preparation and characterization of Hall devices. a** Process flow of making the Hall bar field effect transistor. **b** Photolithography mask of Hall bar field effect transistor. **c** Cross-section structure of Hall bar devices (TG: top gate), (CT: contact). **d** Schematics of the cross-section of the Ge/GeSi heterostructure of the Hall bar device. **e** EDX mapping in the ohmic contact region and **f** the channel region.

Figure 4a shows the four-terminal lock-in measurement circuit. Two synchronous timing lock-in amplifiers allow measurement of the transverse and longitudinal voltages $V_{xy}, V_{xx}$. The transfer characteristic curve is shown in Fig. 4b. It is worth noting that our transfer characteristic curve changes with the scanning range. Holes start to populate the channel at $V_{TH} = -0.8$ V, drift to $V_{TH} = -4.6$ V through multiple scan and saturate at $V_G = -5$ V. The relatively high and unstable threshold voltage, $V_{th} < -4.6$V, is attributed to the negative bias temperature instability (NBTI) effect[27]. The two-dimensional hole carrier density $p_{2D}$ and mobility $\mu$ are obtained by the classical Hall effect. Figure 4c shows the relation between $\mu$ and $p_{2D}$. In the high-density regime, the interface-state density in the silicon-dielectric interface is high and makes it difficult to obtain a stable and large $p_{2D}$; thus, the measurement of the highest mobility under great density is not allowed. While theory indicates a saturated density of $\approx 5 \times 10^{11} \text{cm}^{-2}$, a stable carrier density of $2.195 \times 10^{11} \text{cm}^{-2}$ is chosen, and the mobility under this density is $198.6 \times 10^4 \text{cm}^2/\text{Vs}$, setting new benchmarks for strained Ge devices. We fit the curve to a power law dependence $\mu = p_{2DHG}^{\beta}$ [28]. In the high-density regime, $\beta_{high} = 1.6968$, we can confirm that the majority scattering mechanism is remote charge scattering due to an imperfect silicon-dielectric interface. This finding supports our assumption of a severe NBTI effect. In the low density regime, $\beta_{low} = 10.38$, the nonequilibrium hole tunnels to the surface and leads to surface passivation.[21] A critical density of $p_c = 5.625 \times 10^{10} \text{cm}^{-2}$ is obtained by linear fit $\mu - \ln(p_{2DHG})$, and a percolation density of $p_p = 5.624 \times 10^{10} \text{cm}^{-2}$ is obtained according to percolation theory $\sigma_{xx} \propto (p_{2DHG} - p_p)^p$ (in Fig. 4d, fitting with critical density[29, 30, 31]. The percolation density is lower than in any other strained Ge devices reported with the same GeSi barrier thickness, indicating the low disorder and high quality of the epitaxial material. We note that the curve fitting is in the low-density regime, but the thermal emission process in the ohmic contact introduces extra errors. The error is reduced by the lead probe structure, which has a small ohmic contact resistance because the ohmic contact is operated at a high density area[32]. Figure 4e and Figure 4f show magnetotransport curves, and both the Integer and Fraction Quantum Hall Effect are obvious, proving the high quality of the materials mentioned above.

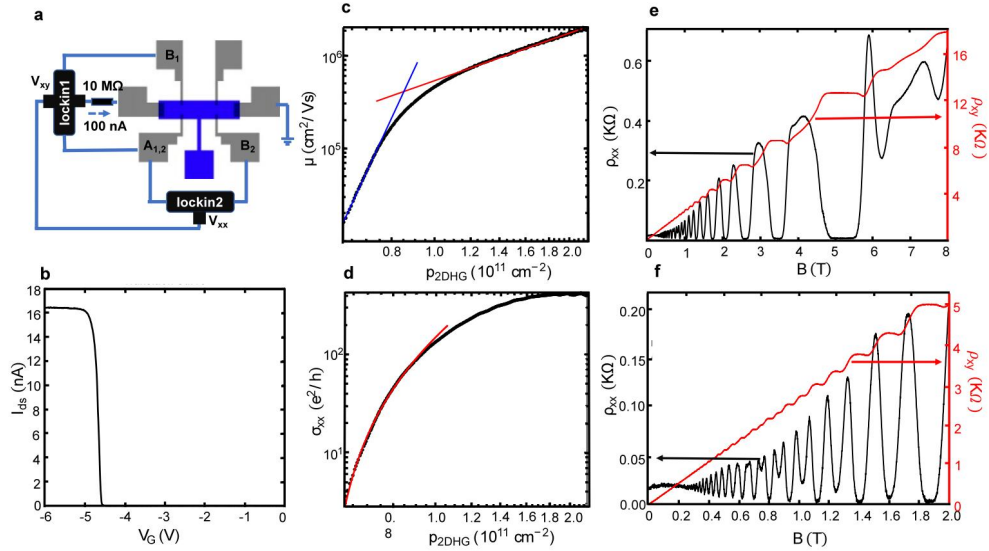

**Fig. 4 Magnetotransport measurements.** Hall-bar shaped HFET. **a** Four-terminal lock-in measurement circuit. The black area is the lock-in amplifier, the green area is the source and drain terminal, and the orange area is the top gate. $V_{sd} = 20$ uV, $f = 230$ Hz. **b** Transfer characteristic curve. **c** Mobility $\mu$ vs density $p_{2DHG}$. Black filled dots are data points. The blue curve is the fitting curve at low density. $\beta_{low} = 10.3824$. The red curve is the fitting curve at high density. $\beta_{high} = 1.6968$. The confidence levels of the two curves are both more than 0.999. **d** Longitudinal conductivity $\sigma_{xx}$ vs density $p_{2DHG}$. Black filled dots are data points. The red curve is fitted by $\sigma_{xx} \propto (p_{2DHG} - p_p)^p$, p=2. The confidence level is 0.999. **e** Quantum Hall effect curve at density $p_{2DHG} = 2.51 \times 10^{11} \text{cm}^{-2}$. **f** Quantum Hall effect curve in the low magnetic field regime at density $p_{2DHG} = 2.71 \times 10^{11} \text{cm}^{-2}$.

To obtain more information, Landau fan diagrams under different temperatures are plotted. Landau fan diagrams of $\rho_{xx}$ and $\rho_{xy}$ vs $p_{2DHG}$ and magnetic field B at 16 mK are shown in Fig. 5a and Fig. 5b, respectively. In the high-density regime, the oscillation begins at $\nu = 22$, which is limited by the measurement accuracy, and $\nu$ is the filling factor. The Zeeman effect appears as $\nu \geq 12$, and an odd filling factor resonates with lines. High mobility allows resonation to start at a low magnetic 0.3T. A Fourier transform is introduced to the SdH oscillation data to calculate the carrier density $p_{SdH}$. The agreement of $p_{SdH}$ and $p_{2DHG}$ in Figure. 5c, the absence of beat frequency and zero longitudinal resistance of the hall platform verify it as a single subband system, i.e., heavy hole system[33]. Figure 5e shows the temperature dependence of the normalized SdH oscillation amplitude at $p_{2DHG} = 2.3 \times 10^{11} \text{cm}^{-2}$. The damping of the amplitude appears as the temperature increases. The hole effective mass $m^*$ is obtained by fitting the damping, as shown in Figure 5f [20, 34]. Similarly, $m^*$ of other densities can be obtained. Figure. 5d shows the m* dependence of $p_{2DHG}$. The black filled dots are extracted data, and the red line is the lineal fitting. According to the fitting, we obtain the effective mass of the top of the valence band $m^* = (0.0728 \pm 0.0066)m_0$ [35]. Next, we try to determine the majority scattering mechanism. The transport lifetime $\tau_t$ is extracted in Figure. 5g and is more than 100 ps, setting a new benchmark. The quantum lifetime $\tau_q$ is obtained from the magnetic damping of the SdH oscillation amplitude, $\tau_q = 1.7 ps$ nearly independent of density fitting to the theory[36]. Figure. 5i shows the Dingle ratio $\tau_t/\tau_q$ vs $p_{2DHG}$, and $\tau_t/\tau_q > 40$ is obtained. A high Dingle ratio, a low $\beta$ in $\mu = p_{2DHG}^{\beta}$ in the high-density regime and the severe NBTI effect together indicate that mobility is limited by the silicon-dielectric interface[37]. The out-of-plane effective g-factor $g^*$ changes with $p_{2DHG}$ in Figure. 5j obtained by Zeeman splitting[38]. The results of low $g^*$ is considered as two reasons. First, as the hole density increases, the Fermi level rises, resulting in a deeper mixture of heavy and light hole subbands. Second, a graded Ge/GeSi interface causes the hole wave function to leak to the GeSi layer and leads to a low $g^*$, which is also attributed to a high $m^*$ [39].

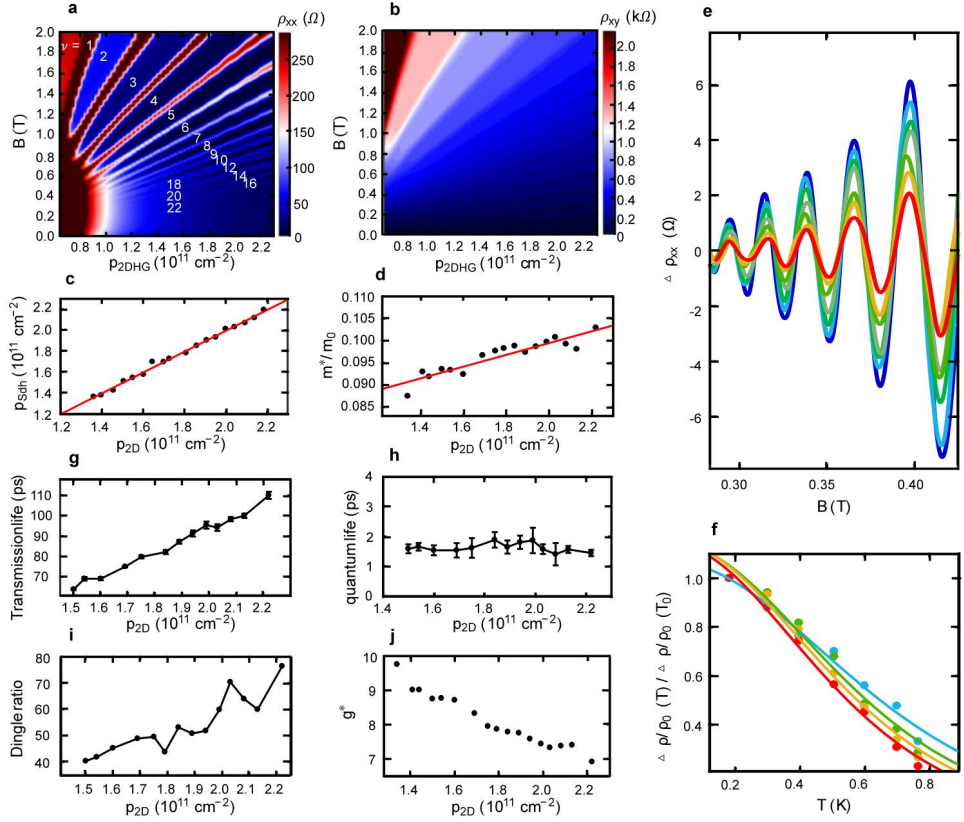

**Fig.5 Magneto transport measurements as a function of temperature, density and magnetism. a** and **b** $T = 16$ mK, $\rho_{xx}$ or $\rho_{xy}$ vs B and $p_{2DHG}$, respectively. **c** $p_{SdH}$ vs $p_{2DHG}$, black filled dots are data, red line is lineal fitting. **d** Effective mass $m^*$ vs $p_{2DHG}$, black filled dots are data, red line is lineal fitting. **e** $\Delta\rho_{xx}$ vs $B$ at $T = 182$ mK $- 766$ mK, $p_{2DHG} = 2.3 \times 10^{11}$ cm$^{-2}$. **f** Temperature damping extracted from fig.e and the fitting curve. **g** Transport lifetime $\tau_q$ vs $p_{2DHG}$. **h** Quantum lifetime $\tau_q$ vs $p_{2DHG}$. **i** Dingle ratio $\tau_t/\tau_q$ vs $p_{2DHG}$. **j** Effective g-factor $g^*$ vs $p_{2DHG}$.

## 3. Discussion

In summary, we prepared a high-quality undoped GeSi/Ge heterojunction with the highest two-million hole mobility currently reported. By fast regrading to 90% Ge content, the penetration dislocation was reduced due to excessive gradients of content. These methods to improve the lattice quality are the key to obtaining high mobility. Simultaneous high lattice quality and carrier mobility imply lower disorder. The 32 nm top barrier is thinner than the reported 1 million migration rate of 66 nm, which is more suitable for the preparation of shallow junction quantum dot devices, and it is easier to serve as a material platform for high-quality quantum dot integrated interconnection. The magnetic transport characterization of Hall-bar shaped H-FETs verifies that the strained germanium system is a low disorder high quality quantum computing platform. The high mobility ($198.6 \times 10^4$ cm$^2$/Vs) and low percolation density ($5.642 \times 10^{10}$ cm$^{-2}$) establish benchmarks for the strained germanium system. However, the device instability, low power exponent and high Dingle ratio in the μ-p relation at high density all indicate that the main scattering mechanism of mobility at high density is remote charge impurity scattering; that is, the impact of the semiconductor/dielectric interface is dominant. In terms of the growth of the wafer, an appropriate thickness of the cap layer should be sought. Second, in the subsequent process preparation, ohmic contact with ion implantation technology can increase the process redundancy of subsequent interface processing compared to the alloy ohmic contact, such as interface cleaning, growth of high-quality ultrathin oxides, and passivation. In conclusion, this research provides a good material basis for the multibit scaling of semiconductor quantum computing.

**Acknowledgements**

This work was supported by Innovation Program for Quantum Science and Technology (Project ID. 2021ZD0302300), National Key research and development plan (Project ID. 2016YFA0301701), the Youth Innovation Promotion Association of CAS (Project ID, 2020037), National Natural Science Foundation of China (Grants No. 61922074 and 12074368, 12034018).

**Author contributions**

Z.Z.K. prepared the heterostructures materials, G.L.W. supervised the material development. Z.H.L. fabricated the Hall bar transistors and measured at low temperature.J.L.S., Y.W.Z., J.B.L., J.X.L., Y.H.R., L.M.W, Y.Y.W., Z.H.W., H.O.L., and J.C.Y. analysed the data;Z.Z.K. and Z.H.L. performed the experiments, supported by G.L.W and G.C. H.H.R., and Y.Y.W., carried out additional analysis. Z.Z.K. and Z.H.L. wrote the manuscript with input from all authors. G.L.W. conceived and supervised the project. C.Z., G.P.G., and J.F.L. provided equipment and environmental support.